# Deep Attention Recognition for Attack Identification in 5G UAV scenarios: Novel Architecture and End-to-End Evaluation

Joseanne Viana∗‡§, Hamed Farkhari∗†§, Pedro Sebastião ∗‡, Luis Miguel Campos †,

Katerina Koutlia ¶, Biljana Bojovic ¶, Sandra Lagén ¶, Rui Dinis‖‡
∗ISCTE – Instituto Universitário de Lisboa, †PDMFC, ‡IT – Instituto de Telecomunicações
‖FCT – Universidade Nova de Lisboa, ¶CTTC - Centre Tecnològic de Telecomunicacions de Catalunya (CERCA);

**Abstract**

Despite the robust security features inherent in the 5G framework, attackers will still discover ways to disrupt 5G unmanned aerial vehicle (UAV) operations and decrease UAV control communication performance in Air-to-Ground (A2G) links. Operating under the assumption that the 5G UAV communications infrastructure will never be entirely secure, we propose Deep Attention Recognition (DAtR) as a solution to identify attacks based on a small deep network embedded in authenticated UAVs. Our proposed solution uses two observable parameters: the Signal-to-Interference-plus-Noise Ratio (SINR) and the Reference Signal Received Power (RSSI) to recognize attacks under Line-of-Sight (LoS), Non-Line-of-Sight (NLoS), and a probabilistic combination of the two conditions. In the tested scenarios, a number of attackers are located in random positions, while their power is varied in each simulation. Moreover, terrestrial users are included in the network to impose additional complexity on attack detection. To improve the system's overall performance in the attack scenarios, we propose complementing the deep network decision with two mechanisms based on data manipulation and majority voting techniques. We compare several performance parameters in our proposed Deep Network. For example, the impact of Long Short-Term-Memory (LSTM) and Attention layers in terms of their overall accuracy, the window size effect, and test the accuracy when only partial data is available in the training process. Finally, we benchmark our deep network with six widely used classifiers regarding classification accuracy. Our algorithm's accuracy exceeds 4% compared with the eXtreme Gradient Boosting (XGB) classifier in LoS condition and around 3% in the short distance NLoS condition. Considering the proposed deep network, all other classifiers present lower accuracy than XGB.

## I. Introduction

unmanned aerial vehicles (UAVs) have the potential to bring revolutionary changes that will fulfill consumer demands in several industry verticals. UAVs will play a crucial role in emergency response [1, 2], package delivery in the logistics industry, and in temporal events, [2]. UAVs are becoming more common and reliable [3] due to technological advancements [4, 5], as well as the improvements in energy-efficient UAV's trajectory optimizations algorithms to be feasible in practice to take into account the dynamics of the UAV as a parametrized method [6, 7, 8], thus integrating UAVs into 5G and 6G networks will increase telecommunication coverage and reduce costs for businesses willing to invest in this technology.

However, UAVs can easily be hacked by malicious users [9] throughout their wireless communication channels, which might divert delivery packets from their destinations. This can have disastrous consequences in unfortunate climate events where UAVs are transporting people to hospitals, or in cases of criminal investigations. A jamming attack can lead to loss of UAV communication control, UAV robbery, UAV destruction, and property damage in urban areas, which would generate problems for business leaders. The authors in [10, 11, 12, 13], emphasize the need for research on new robust methods for attack detection and its associated challenges in 5G UAV communications.

Obviously, the ability to recognize different patterns in communication connectivity plays an important role in the UAV security paradigm. Therefore, a Self-Identifying Solution against Attacks (SISA), becomes a basic requirement for UAV communication control. According to [14], identification of interference must serve as the basis for selecting anti-jamming solutions. Statistical models have recently been recognized as a viable way for monitoring network activity in wireless communications and detecting suspicious attacks through the use of wireless parameters. Cheng et al. [15] offer a Bayesian technique for detecting jamming. The authors of [16] present a jamming detection approach based on a Naive Bayes classifier trained on a small sample of data and addresses just noise effects. The authors in [17] employ a sequential change point detection algorithm to detect the state changes in the time series using Bayesian estimators. Lu et al. [18] propose the message invalidation ratio as a new metric to evaluate performance while under jamming attacks in time-critical applications. In [19], the authors offer a jamming detection strategy for Global Navigation Satellite System (GNSS)-based trained localization that makes use of Singular Value Decomposition (SVD). However, the majority of research does not account for the effects of the wireless propagation channel in their solutions.

§Collaborative authors with equal contribution

TABLE I: Abbreviation list.

| Abbreviation | Definition | Abbreviation | Definition |
|---|---|---|---|
| ASD | Azimuth spread of departure | LR | Logistic Regression |
| ASA | Azimuth spread of arrival | LSTM | Long Short-Term Memory |
| A2G | Air to ground | MVA | Majority Voting Algorithm |
| CAT | CatBoost | NLoS | Non-Line-of-Sight |
| CDL | Clustered Delay Line | RF | Random Forest |
| CNN | Convolutional Neural Network | SINR | Signal-to-Interference-plus-Noise Ratio |
| 1 CPU | Central processing unit | SISA | Self-Identifying Solution against Attacks |
| C-RAN | Cloud Radio Access Network | SVD | Singular Value Decomposition |
| DAtR | Deep Attention Recognition | SVM | Support Vector Machines |
| DL | Deep Learning | RSSI | Reference Signal Received Power |
| DNN | Deep Neural Networks | TSA | Time-Series Augmentation |
| GNB | Gaussian Naive Bayes | UAV | Unmanned Aerial Vehicle |
| GNSS | Global Navigation Satellite System | UMi | Urban Micro Scenario |
| MH-DNN | Multi-Headed Deep Neural Network | XGB | eXtreme Gradient Boosting |
| ML-IDS | Machine learning Intrusion Detection System | ZSD | Zenith spread of departure |
| LoS | Line-of-Sight | ZSA | Zenith spread of arrival |

With respect to machine learning, Krayani et al. use a Bayesian network to identify jammers in [20]. Youness et al. [21] create a dataset based on signal property observations and use Random Forest (RF), Support Vector Machines (SVM), and a neural network algorithm to classify the features extracted by the jamming signal. [22] also uses a SVM and a self-taught learning method to identify attacks in UAV Networks. In [23], the authors utilize a Machine Learning Intrusion Detection System (ML-IDS) based on SVM to identify jamming in the Cloud Radio Access Network (C-RAN). Deep Learning (DL) has been used to create models with high-level data abstraction by utilizing numerous layers with activation function processing. In DL, Deep Neural Networks (DNNs), such as Convolutional Neural Network (CNNs) [24, 25] are able to define trends and seasonality in time-series data [25, 26]. These characteristics make deep network-based algorithms useful for discovering patterns in wireless networks by analyzing time series and spatial information [27]. The authors in [28] also identify jamming samples using signal-extracted features, but the authors add another way to detect attacks that employs 2D samples and pre-trained networks, such as AlexNet, VGG-16, and ResNet-50. In [29], the authors also use pre-trained deep networks to develop a three step framework to identify jamming in radar scenarios. In [30], the features of the signal in the time domain, frequency domain, and fractal dimensions, as well as deep networks, are used to recognize jamming attacks.

Nevertheless, Deep Learning (DL) presents its own challenges when applied in the wireless context:

- It is challenging to collect network parameters for DL input layers. All deep learning algorithms need training and testing. In each phase, the DNN's input layer is made up of the parameters of the data samples. The greater the sample coverage in terms of data qualities, the better the DL can identify network features. However, some wireless data may be missing due to the stochastic nature of the communication paths. As a consequence, DL models should be built to tolerate missing parameters, data errors, and out-of-range values in their input layers.
- UAVs have constraints in memory, CPU capabilities, and available batteries. Complex algorithms cannot be programmed into their current protocols because DL is iterative in nature. This may prolong system response time. To save memory space, the DL algorithms should use techniques that do not rely on increasing the amount of layers, nodes, or trainable parameters. To minimize execution time, the algorithms should be optimized.
- DL needs entire or nearly complete training samples to effectively detect network patterns. However, because of the difficulty of collecting so many data points for each potential network condition, the training samples may be relatively restricted. This dictates that DL should be capable of adding additional samples after failing to recognize a new pattern. The fresh samples may help to increase the accuracy of the DL models.
- Furthermore, network engineers/programmers are required to carefully design the DL data formats since various network parameters have extremely distinct data properties and formatting requirements. The correct numerical representations and data normalization algorithms must be explicitly stated to combine numerous network parameters into the same DL input layer.

### *A. Objectives and contributions*

In this paper, we study the attack identification problem in authenticated UAVs in 5G communications. To enable UAVs to cope with jamming recognition, we propose a deep network called DAtR (Deep Attention Recognition) that uses only two observable parameters: Signal-to-Interference-plus-Noise Ratio (SINR) and Reference Signal Received Power (RSSI). 5G communication networks provide these measurements in the receivers in LoS conditions. We add NLoS, and probabilistic LoS and NLoS conditions in the deep network and compare the accuracy for each channel condition case. We use a neural network that includes attention layers with optimized parameters to decrease the chances of low accuracy when adding users and attackers to the network. We demonstrate that the DAtR is able to recognize jamming attacks from other malicious aerial agents in complex urban environments, where there are terrestrial users connected to the network. The final goal is to demonstrate that it is possible to identify attacks in the UAV's receiver using learning techniques, such as deep network architectures, which have significantly fewer layers than well-known pre-trained networks. Also, the deep network does not rely on transfer learning techniques, and it could provide better accuracy than other well-known classifiers.

Taking these into account, the main contributions of this work are highlighted in the following:

1) A novel, robust, and effective convolutional-attention deep network for UAVs, named DAtR, that detects jamming in complex environments under LoS and NLoS conditions and that tolerates incomplete raw data inputs. To the best of the authors' knowledge this is the first time that an attention model is proposed to detect jamming in LoS, NLoS, and hybrid conditions.
2) A study of deep network architectures for UAVs considering Long Short-Term Memory (LSTM) and Attention layers for 5G UAV communication data.
3) Two new complementary methods named Time-Series Augmentation (TSA) and Majority Voting Algorithm (MVA) to improve classification accuracy and detect false alarms for deep networks.
4) An accuracy comparison with six other state-of-the-art machine learning classifiers.
5) An analysis of the tradeoffs between accuracy and added latency in the model while identifying attacks.

The remaining parts of this paper are organized as follows. Section II presents the preliminaries and the attack identification problem in authenticated UAVs. Additionally, it describes the transmission and channel models, as well as the observable parameters of SINR and RSSI. Section III illustrates the proposed deep network for jamming identification. Section IV focuses on the accuracy analysis of the network simulation results, comparisons of parameter configurations, and comparisons between the proposed deep network with six different classifiers. Section V includes our conclusions. Table I summarizes the abbreviations used in this paper.

## II. Preliminaries and Problem Formulation

### *A. Scenarios*

Fig. 1 illustrates the UAV simulation environment. It identifies the adopted X-Y-Z cartesian coordinates. We consider a scenario where authenticated UAVs fly in a 1 km × 1 km square area, while they are connected to a serving small cell through Air-to-Ground (A2G) 5G wireless data links. In this environment, we include authenticated terrestrial users placed on the ground. UAV attackers are placed in predetermined randomly assigned spots. They fly towards the authenticated UAVs inside the coverage area of the small cell. To create our model, we assume that the authenticated UAV transmission power is fixed during each simulation, and we use Clustered Delay Line (CDL) channels including slow and fast fading components to model their propagation conditions. UAV attackers use the same propagation models as the authenticated UAVs [31],[32]. For the terrestrial users, we follow the 5G wireless terrestrial propagation models defined in [32] instead. Fig. 1 shows a configuration example with two authenticated UAVs, three terrestrial users, three UAV attackers, and one small cell.

For the sake of simplicity, the authors considered the UAV to be a fly antenna; Assuming that the antennas position in the UAV is ideal, that means their present the best performance possible to the simulation results and the UAV sizes or UAVs mechanical components have a low contribution to the overall experiment.

When UAV attackers move, their speed is kept constant, and they head in the direction of the authenticated UAVs getting closer to them as simulation time evolves. The attackers' and authenticated UAVs' positions are at higher altitudes and follow the losses according to the standards in [31] and [32]. Our research presumes that terrestrial users may likewise be in fixed locations or have the ability to change their positions according to mobility models in [33]. The small cells are configured with an antenna height of 10 m typically seen in the urban environments. Table II displays the four different experimental setups we created, in which basically multiple combinations of mobility for UAV attackers and/or terrestrial users are considered. During the simulations, as further explained in Section V, we vary the scenarios to account for different mobility/speed options, as well as different distances between the small cells and authenticated UAVs, UAV attacker power, number of UAV attackers, and number of terrestrial users.

The authenticated UAVs try to identify if there are any attackers attempting to disrupt the communication link by using the proposed DAtR mechanism, which is fed with the RSSI and SINR measurements that are available in the receiver. For each scenario listed in Table II, we create a dataset with 600 files, including up to four attackers and thirty terrestrial users connected at the same time. We group them together to form a complete dataset, composed of 2400 files split into RSSI and SINR parameters in constant LoS condition. Then, we change the channel condition in the dataset and check if it is possible to identify the attackers in persistent NLoS condition, and in randomly combined LoS and NLoS conditions through the 3rd Generation Partnership Project (3GPP) stochastic models in [31] and [32]. In the end, we have 3 datasets with 2400 files each, corresponding to LoS, NLoS, and hybrid LoS/NLoS conditions. Additional information on the dataset's development and possible applications are available in [34, 35]. This is an intriguing problem due to the fact that in LoS cases, channel variations and terrestrial users increase the difficulty of identifying attacks. Under the NLoS condition, the lower received power makes it more challenging to recognize the UAV attackers. Finally, let us notice that the connection link between the authenticated UAV and the small cell exists during the entire simulation even in low SINR circumstances.

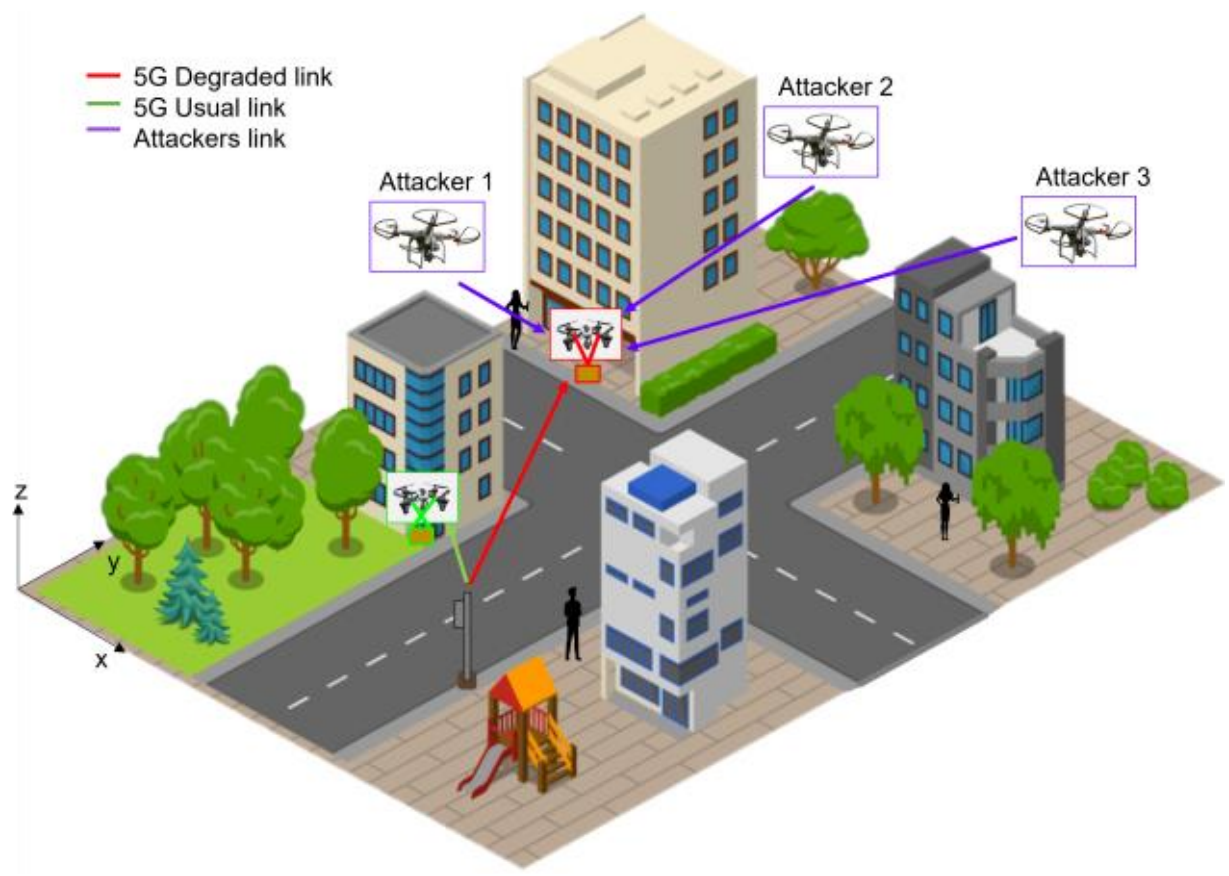

Fig. 1: Simulation scenario.

TABLE II: Speed configuration scenario.

| Scenario | Attackers configured with speed | Users configured with speed |
|---|---|---|
| None Speed | N | N |
| Attackers Speed | Y | N |
| Users Speed | N | Y |
| Both Speed | Y | Y |

### *B. Communication model*

We consider an A2G connection between the small cell and the authenticated UAVs, as depicted in Fig. 1. The scenario consists of an urban environment where buildings, trees, and other structures may cause significant path loss and shadowing degradation. We define the A2G large scale effect with two components, i.e., path loss and shadowing, as follows:

$$L^{a}(d, f) = PL^{a}(d, f) + \eta^{a} \text{ [dB]}, \tag{1}$$

where $PL^{a}(d, f)$ is the path loss at distance $d$ from the authenticated UAV to the respective small cell (in km), when transmitting over the carrier frequency $f$ (in MHz). $\eta^{a}$ is the shadowing (in dB), and $a$ reflects the LoS and NLoS conditions, i.e., $a \in \{\text{LoS}, \text{NLoS}\}$.

In A2G communications, the pathloss $PL^{a}(d, f)$ in Eq. (1) depends on the high/low altitude configurations and the LoS/NLoS conditions. We compute it as follows:

$$PL^{a}(d, f) = \begin{cases} PL^{\text{LoS}}(d, f) & \textit{if } \text{LoS} \\ PL^{\text{NLoS}}(d, f) & \textit{if } \text{NLoS}. \end{cases} \tag{2}$$

For urban UAV scenarios, the pathloss in the LoS condition is given by the maximum between high/low altitude pathloss computations:

$$\begin{aligned} PL^{\text{LoS}}(d, f) &= \max(PL_{h}(d, f), PL_{l}(d, f)). \\ PL_{h}(d, f) &= 20\log(d) + 20\log(f) + 20 log(4\pi/c) \\ PL_{l}(d, f) &= 30.9 + (22.25 - 0.5\log(h))\log(d) + 20\log(f) \end{aligned} \tag{3}$$

where $c$ is the speed of light (in m/s), $h$ is the altitude (in m), $PL_{h}(d, f)$ is the free space path loss for high altitudes, and $PL_{l}(d, f)$ is the low altitude path loss.

Under NLoS condition, the pathloss is given by the maximum between the LoS pathloss and the NLoS pathloss expression:

$$\begin{aligned} PL^{\text{NLoS}}(d, f) &= \max(PL^{\text{LoS}}(d, f), PL_{n}(d, f)) \\ PL_{n}(d, f)_{a} &= 32.4 + (43.2 - 7.6\log(h))\log(d) + 20\log(f) \end{aligned} \tag{4}$$

In our scenario, we assume that all the UAVs fly with a height within the margin 22.5m < $h$ < 300m. With that in mind, the remaining shadowing component ($\eta^a$) in Eq. (1) is defined by 3GPP as an additional variation over the pathloss with a certain standard deviation, depending on LoS/NLoS conditions as well. Table III includes the shadowing characterization for LoS and NLoS.

TABLE III: Shadowing for UAVs in UMi [32, 31].

| | Std. deviation (dB) | Altitude (m) |
|---|---|---|
| LoS | $\max(5 \times \exp(-0.01h), 2)$ | $22.5 < h < 300$ |
| NLoS | 8 | $22.5 < h < 300$ |

To determine the LoS or NLoS condition for each communication link, 3GPP uses a stochastic model. The probability of being in LoS ($p_{\mathrm{LoS}}$) is given by:

$$p_{\mathrm{LoS}} = \frac{d_1}{d_{2D}} + \exp\left(\frac{-d_{2D}}{p}\right)\left(1 - \frac{d_1}{d_{2D}}\right), \tag{5}$$

where $p = -233.98 \log_{10}(h) - 0.95$, $h$ is the height of the UAV, $d_1 = \max(294.05 log_{20}(h) - 432.94, 18)$, and $d_{2D}$ is the 2D distance between the UAV and the small cell. Accordingly, the probability of being in NLoS is: $p_{\mathrm{NLoS}} = 1 - p_{\mathrm{LoS}}$.

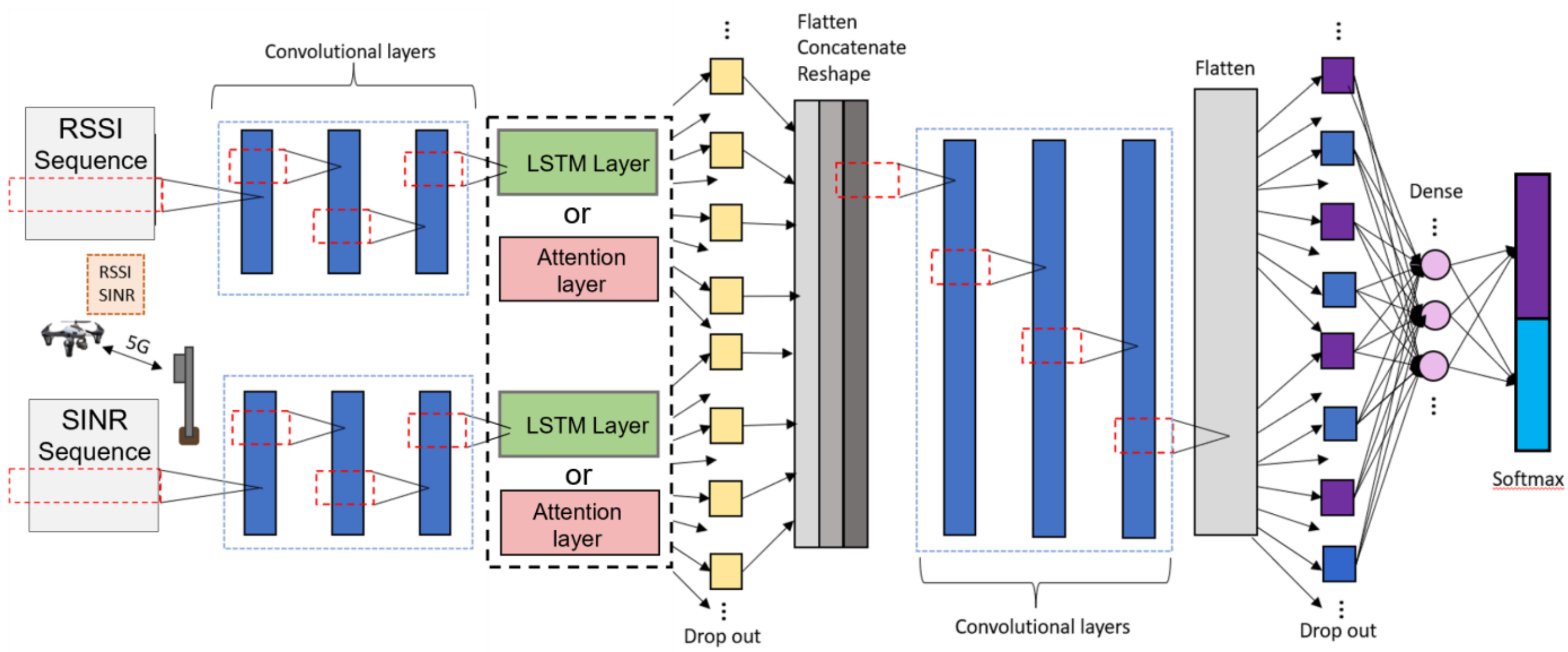

Fig. 2: Multi-headed deep neural network (MH-DNN) architecture. Note the switch from LSTM to Attention layers.

For the small-scale fading, we adopt Cluster-Delay-Line (CDL) models, as in [32] and [31]. 3GPP defines in tabular mode the parameters that model the fading, including the powers, delays and angles of arrival and departure (AoA, AoD) that contain spreads in both the azimuth (ASA, ASD) and zenith (ZSA, ZSD) of each cluster for the UAV scenario. The scenario assumes large and small-scale fading in the link between the UAVs and the small cells.

Given this model, the received power at the UAV with no jammers or interferences can be expressed as:

$$P_{uav} = P + G - L^a(d, f) - S(n, m) \tag{6}$$

where $P$ is the transmission power, $G$ is the overall antenna gain in the link considering UAV and small cell antenna gains, i.e., $G = (G_{uav} + G_{sc})$, and $S(n, m)$ is the small-scale fading effect, which corresponds to the superposition of $n$ clusters with $m$ rays in the communication link, as per [32, 31]. Our model considers single antenna elements in the small cell and the UAVs. The simulation in this work uses CDL-A and CDL-D models for small-scale fading in the NLoS and LoS conditions. In this case, each CDL comprises 23 clusters with 20 multipath components (rays) each. Each cluster has an AoD and an AoA. These values are used to create the rays' AoAs/AoDs according to the azimuth/zenith arrival/departure spreads (ASA/ASD, ZSA/ZSD), respectively.

The signal-to-interference noise ratio (SINR) between the authenticated UAV and the small cell at distance $d$, in the presence of interference coming from jammers and terrestrial users, is given by:

$$\Gamma_{uav} = \frac{P_{uav}}{\zeta^2 + \sum_{i=1}^{U} P_{\text{user}}^{i} + \sum_{j=1}^{J} P_{\text{jammer}}^{j}}, \tag{7}$$

where $P_{\text{user}}^{i}$ and $P_{\text{jammer}}^{j}$ represent the received power at the UAV coming from the i-th user and the j-th jammer, respectively, which act as interfering signals (including the channel gain with the authenticated UAV), $\zeta^2$ is the noise power, $U$ is the total number of terrestrial users transmitting at the same time as the authenticated UAV, and $J$ is the number of jammers transmitting in the scenario.

Next, the RSSI includes the linear average of the total received power in Watts from all sources, including co-channel serving and non-serving cells, adjacent channel interference, thermal noise, etc. Considering $\wedge_0$ as the RSSI value at a reference distance, we have

$$\wedge = \wedge_0 - 10\rho\log(d), \tag{8}$$

where $\rho = L^{a}(d, f) + S(n, m)$ includes path loss and fast fading components and $d$ is the link distance.

### C. Problem formulation and dataset

The SISA goal for the authenticated UAV is to quickly identify malicious changes in the received power, caused by UAV jammers in the environment. For that, we use a small deep network, where the number of trainable parameters T are smaller than 100K ($T < 100000$), that is composed of a combination of layers, including CNNs, Attention, Drop out, Flatten, among others. The details of the DNN architecture are provided in Section III.

First, we study the case where UAV attackers try to disrupt the communication when the UAV and the small cell can directly see each other (LoS condition). Then, we simulate the NLoS condition, where buildings and other elements in the city may block the direct communication between the UAV and the small cell. Finally, we study a probabilistic combination of LoS and NLoS conditions. As such, we assume the following in the three datasets we create for the experiment:

- LoS: The UAV is always in LoS condition throughout all the simulations available in the dataset;
- NLoS: The UAV is in NLoS condition for the entire time during all the simulations included in the dataset;
- LoS and NLoS: The link between the UAV and the small cell is in either LoS or NLoS condition with a probability of $p_{\text{NLoS}}$ and $1 - p_{\text{LoS}}$ (according to Eq. (5)) for all the simulations in the dataset.

Table II describes the four scenarios in each dataset. The differences between the scenarios inside the dataset relates to the following parameters: the UAVs' and terrestrial users' mobility and speed, the distance between the small cell and the authenticated UAVs, the number of attackers and their power, and the number of terrestrial users in the network. It is important to note that the scenarios in the dataset, such as `Attackers Speed`, `Users Speed`, `Both speed` and `None Speed`, are unbalanced, meaning that the proportion between attackers and no attackers in the raw data is different. For example, the dataset has data for 1, 2, 3, 4 attackers, while for no attacks there is 0 attacker data. Therefore, in order to avoid bias towards the classification, it is necessary to implement countermeasures to balance the data during the pre-processing phase.

Our deep network design aims to achieve maximum performance. To this end, we compare the use of LSTM and Attention layers. We improve the capabilities of the Multi-Headed Deep Neural Network (MH-DNN) by integrating TSA and MVA techniques, which results in the proposed DAtR. We benchmark our DAtR with six other well known ML algorithms and analyze other parameters, such as the optimum window size, the attack accuracy when the deep network sees the data for the first time during the test, and the latency added due to the DAtR processing time.

## III. Convolutional-Attention-Based Attack Detection

Our SISA model is based on a MH-DNN. The proposed architecture is shown in Fig. 2. It contains (i) three CNN layers, (ii) an Attention or an LSTM layer, and (iii) a drop-layer in each head. The body of the deep network consists of: (i) a Flatten, (ii) a Concatenate, (iii) a Reshape, (iv) three CNN layers, (v) a Flatten, (vi) a Drop out layer, (vii) a Fully connected layer, and (viii) the output layer for two classes classification. Although RSSI and SINR measure different parameters from the telecommunication perspective, both values may be related to each other. For example, when RSSI increases, SINR may decrease. Using our proposed MH-DNN, we can extract features from both parameters simultaneously in each head at each window size. The window size defines the amount of data the deep network algorithm will receive as input in each head.

First, we convolute both signals in each head in the three CNN layers as Fig.2 indicates. This operation creates a convolution kernel that is convoluted with the layer input over a single temporal dimension to produce a tensor of outputs. Thanks to the configuration of strides and kernels, this operation returns a single vector with several channels (i.e., $1 \times channel$). The convolution operation extracts different features from the time-series data available in each head. The result from the convolutional layers is computed in parallel in the Attention layer, and it is possible to check the states before the current state in the window sequence to understand the attack pattern. The layer uses an auxiliary vector that stores the previous hidden states and increases or decreases the weights of the layer by the sum of the row vectors that hold the information using both the previous and the current states [36]. In our case, we use 8 heads in the Attention layer to capture different contextual information.

The drop out layer removes some of the features from the attention layer to avoid over-fitting. In other words, it prevents the deep network from memorizing the input parameters instead of learning the patterns in the sequences. In the deep network body illustrated in Fig. 2, after the dropout layer, the remaining features from RSSI and SINR are flattened, concatenated, and reshaped.

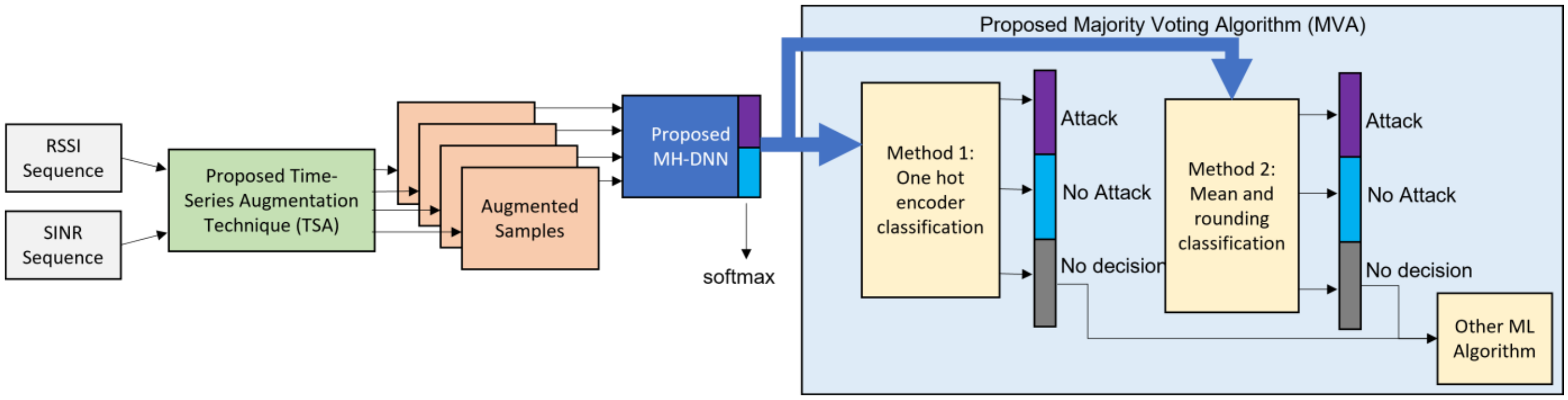

Fig. 3: Deep Attention Recognition (DAtR), including TSA and MVA Techniques - Method 1 and 2

TABLE IV: Deep Network Configuration Parameters.

| Deep network Parameters | Value |
|---|---|
| Base learning rate | $2.5\text{x}10^{-2}$ |
| Base batch size | 32 |
| —— Head —— | |
| Conv-1 filters, kernel size, strides | 8, 8, 2 |
| Conv-2 filters, kernel size, strides | 8, 4, 2 |
| Conv-3 filters, kernel size, strides | 8, 3, 1 |
| Self-Attention head-number, key-dimensions | 8, 8 |
| (or LSTM) | 16 |
| —— Body —— | |
| Conv-1 filters, kernel size, strides | 8, 8, 2 |
| Conv-2 filters, kernel size, strides | 8, 4, 2 |
| Conv-3 filters, kernel size, strides | 8, 3, 1 |
| Fully connected layer | 100 |
| Drop-out | 0.4 |
| Softmax | 2 |

The concatenation procedure merges the features extracted from both RSSI and SINR in each head, while the flatten and reshape methods keep the tensor size consistency. Next, we use the 3-CNN layers again, we invert the flatten with the drop out layer position. Then, we apply dense and softmax layers. The classification happens in the softmax layer after feature extraction and learning representations of the input data. A fair comparison between LSTM and the Attention layer's overall performance requires that the input and the output of both layers have the same size and shape. In order to guarantee that, we add the global average after the Attention layer and we adopt LSTM with 16 filters. Table IV shows the main parameters for the deep network.

## IV. Improvements in MH-DNN Robustness

In this section, we introduce the TSA method combined with the MVA to improve the performance of our deep neural network under the NLoS condition, which tends to present lower total received power compared to the LoS condition. Fig. 3 summarizes the main additions to the MH-DNN in order to include these two new features. Notice that the MH-DNN combined with the TSA and MVA results to the proposed DAtR.

### A. *Time-series augmentation technique*

TSA aims to supplement the original dataset with additional and unrelated samples for the MH-DNN to process further. We create the additional data using data augmentation and flipping techniques that are applied in the train set to increase data diversity, to prevent over-fitting in the test set, and to convert binary classification into 3 classes in the majority voting calculation in section IV-B. As Fig. 3 shows, we convert the input samples into four augmented samples. In Table V, we display an example of how to generate the four new augmented samples according to TSA.

By randomly inverting each RSSI and SINR sequence, we are able to generate four distinct augmented samples from each occurrence. Let us point out, that other data augmentation strategies could be considered to generate the data as well. After

TABLE V: Output of the TSA.

| | RSSI Sequence | SINR Sequence |
|---|---|---|
| Sample 1 | Same | Same |
| Sample 2 | Same | Flipped |
| Sample 3 | Flipped | Same |
| Sample 4 | Flipped | Flipped |

**Algorithm 1** Majority Voting Algorithm

**Require:** $\tau, T$
**Ensure:** Assign $\tau$ to Classes 1 or 2 or 3
  $Class$ 1 || $Class$ 2 ← $Classify$ T
**if** $3T/4 \geq Class$ 1 **then**
  $Class$ 1 ← $Classify$ $\tau$
**else if** $3T/4 \geq Class$ 2 **then**
  $Class$ 2 ← $Classify$ $\tau$

**else if** $T/2 == Class$ 1 and another $T/2 == Class$ 2 **then**
  $Class$ 3 ← $Classify$ $\tau$
**end if**

preprocessing the dataset, which results in the conversion of the data to augmented samples with an appropriate rolling window, each augmented sample has two data sequences representing the RSSI and the SINR. Then, we feed the augmented samples to MH-DNN as in Fig. 3.

### *B. Proposed majority voting algorithm*

DAtR uses TSA and MVA as preprocessing and postprocessing techniques, respectively. After, feature classification is done in the softmax layer, we use the MVA to reclassify the features in order to have better accuracy.

MVA is divided into 2 Methods (see Fig. 3). Initially, in Method 1, MVA uses one hot encoding probability values between 0 and 1 as input from the MH-DNN classification prediction, rounds them, then calculates the mean for each sample created from the previously explained TSA method, and uses one hot encoding to classify it again. If the feature is classified in class 1 (attack) or 2 (no attack), the code finishes and the classification achieves high accuracy, minimal false alarms, and the amount of features in class 3 (no decision) is low. However, if the feature is classified in class 3, we try to reclassify using other ML algorithms. In Method 2, we try to classify the features as class 1 or 2 by inverting the algorithm order. Instead of rounding it first and then calculating the mean, we calculate the mean and then round it. The mean calculation at first decreases the precision of the encoding features and consequently the overall accuracy, which increases the chances of a false alarm and includes a higher degree of unclassified data (ud). If after Method 2, the feature can not be classified in class 1 or 2, we apply other well known ML algorithms to classify the features that Methods 1 and 2 could not classify. Notice that although the proposed DAtR results to be efficient in LoS channel conditions (as it will be demonstrated in Section X), the motivation for using preprocessing and postprocessing techniques in MH-DNN arises from the fact that the attack detection accuracy might decreases in cases of extremely low received power conditions, as they happen in NLoS channel conditions. As such, we target to increase accuracy by applying TSA and MVA. At the end, DAtR proved to be efficient also in LoS conditions.

Algorithm 1 illustrates the details of Methods 1 and 2 where $\tau$ is the main sample and $T$ represents the four augmented samples for the $\tau$ sample. The algorithm tries to classify the deep network features when the categorization into the classes is not possible in the softmax layer. Basically, if 3 out of the 4 augmented samples classify a feature into a specific class, then this feature is added to the class. In the case of a draw, the feature goes into class 3.

## V. Simulation Results

In this section, we present the performance evaluation of the proposed DAtR. In particular, we provide five experimental outcomes related to the robustness of the DAtR. As a first step, we conduct a comparative study on the efficacy of different layers, such as Attention and LSTM in the MH-DNN architecture. Then, we study the effect of the window size on the DAtR's accuracy. In addition, we examine the performance of the proposed DAtR when we remove parts of the dataset from training, and we benchmark the DAtR's accuracy against six machine learning alternatives. All these experiments evaluate LoS and NLoS channel conditions, separately. For the evaluation of the DAtR's performance, we compare the overall accuracy based on the various parameters available in the dataset. Initially, we analyze the accuracy as a function of the number of attackers and attackers' power. After that, we analyze the accuracy as a function of the attackers' distance and power. These simulations set all the three conditions presented in the paper side by side; LoS, NLoS, and a combination of LoS/NLoS, respectively.

Table VI presents the parameters used in the simulation. The speed remains the same for all scenarios and the distances in Table VI refer to the distances between the small cell and authenticated UAVs.

TABLE VI: Network Parameters.

| Scenario Parameters | Value |
|---|---|
| Terrestrial Users | 0, 3, 5, 10, 20, 30 |
| Authenticated UAVs | 1 |
| Small Cells | 10 |
| Small cell height | 10 m |
| Attackers | 0, 1, 2, 3, 4 |
| Speeds | 10 m/s |
| Modulation scheme | OFDM |
| Small cell power | 4 dBm |
| Authenticated UAV power | 2 dBm |
| Attackers power | 0,2,5,10,20 dBm |
| Authenticated UAV position | URD* |
| Attackers position | URD* |
| Small cells position | URD* |
| Scenario | UMi |
| Distance | 100, 200, 500, 1000 m |
| Simulation time | 30 s |

*URD - Uniformly Random Distributed

### A. *The window size impact*

Fig. 4 (a) and (b) show the window size impact on the final accuracy for LoS and NLoS conditions, using the MH-DNN (no improvements, no TSA and no MVA). Fig. 4 (a) indicates that the accuracy range for w = 100 is roughly 65% to 90%, whereas the range for w =300 is approximately 75% to 95%. In the NLoS case (see Fig. 4 (b)), the MH-DNN achieves a range of about 67% to 85 % when w = 100, and the percentage ranges from 70% to 87 % when w = 300.

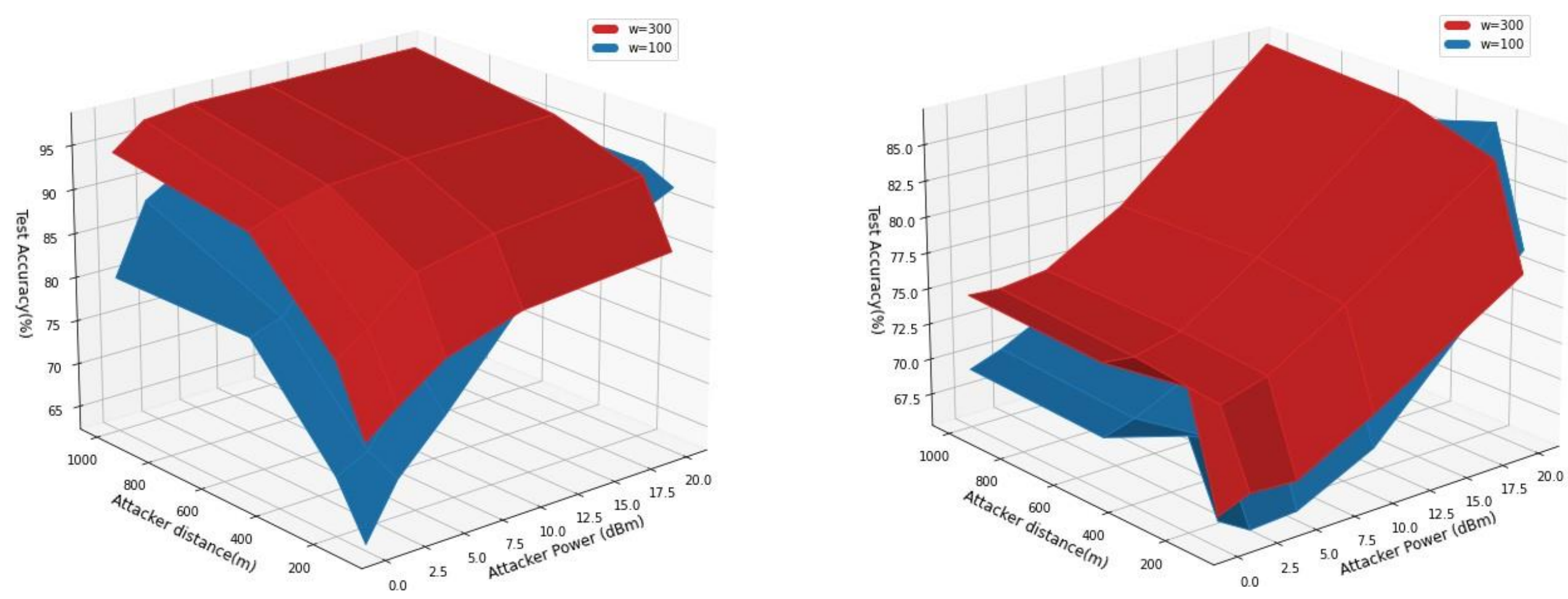

(a) LoS Condition

(b) NLoS Condition

Fig. 4: Impact of the window size w=100 and w=300 (a) In LoS, (b) In NLoS

Both figures demonstrate that the accuracy is directly proportional to the window size, independently of the channel condition. It is worth noting that there is a small tradeoff between the time that it takes to calculate the estimate for each class and the available resources (i.e., the window size), as it will be demonstrated later in Fig. 11.

### B. *Attention vs LSTM*

Both the LSTM and Attention layers are trying to solve the same problem. They keep track of the old input sequences in the current node or state. For example, the information flowing from $t_0$ to $(t - n)$ is available in a modified/partial form in the state at time $t$. The algorithm uses the modified form to establish a relationship with the incoming data. We opt to compare both LSTM and Attention in terms of window size and final accuracy improvements in LoS and NLoS conditions for each proposed algorithm in the paper. The trainable parameters do not change between the different window sizes or conditions. In our example, the MH-DNN configured with LSTM has 59984 trainable parameters compared to 64368 in the one with the Attention. The majority of well-known pre-trained deep neural networks, such as VGG [37] and ResNet [38], employ more than one million trainable parameters in their architectures, which increases the overall training time and requires more computation capabilities. During the test, we only interchange the Attention and LSTM layers, using the settings in Table IV and the proposed DAtR.

TABLE VII: Differences in the overall accuracy for each condition and for each window size

| w | | | 50 | 100 | 200 | 300 |
|---|---|---|---|---|---|---|
| DNN | LoS | **Attention** | **82.26** | 83.04 | **88.35** | **89.59** |
| | | LSTM | 79.62 | **84.67** | 86.51 | 88.06 |
| | NLoS | **Attention** | **72.58** | **73.00** | **74.12** | **75.60** |
| | | LSTM | 69.43 | 71.46 | 65.76 | 68.67 |
| | Both | **Attention** | **76.31** | **79.59** | **79.19** | **82.77** |
| | | LSTM | 76.07 | 78.19 | 77.10 | 77.29 |
| DNN+Method 1 | LoS | **Attention** | **83.88** | 84.31 | **88.48** | **89.98** |
| | | LSTM | 83.65 | **84.38** | 87.10 | 88.34 |
| | NLoS | **Attention** | **82.81** | 82.53 | **82.94** | **83.07** |
| | | LSTM | 81.87 | **83.05** | 81.27 | 80.19 |
| | Both | **Attention** | **80.50** | **81.27** | **79.13** | **83.66** |
| | | LSTM | 79.82 | 79.67 | 78.95 | 79.02 |
| DNN+Method 2 | LoS | **Attention** | **84.10** | 84.77 | **89.99** | **90.80** |
| | | LSTM | 81.34 | **86.26** | 88.47 | 89.49 |
| | NLoS | **Attention** | **75.66** | **76.07** | **77.13** | **79.00** |
| | | LSTM | 72.20 | 73.85 | 68.60 | 73.10 |
| | Both | **Attention** | **78.61** | **81.52** | **80.51** | **84.65** |
| | | LSTM | 78.28 | 80.11 | 79.22 | 79.59 |

TABLE VIII: Accuracy measurements using XGB for each condition and for each window size

| XGB | w=50 | w=100 | w=200 | w=300 |
|---|---|---|---|---|
| LoS | 83.27 | 83.69 | 85,57 | 86.33 |
| NLoS | 83.04 | 82.58 | 83.41 | 80.58 |
| Both | 79.65 | 79.47 | 78.40 | 78.85 |

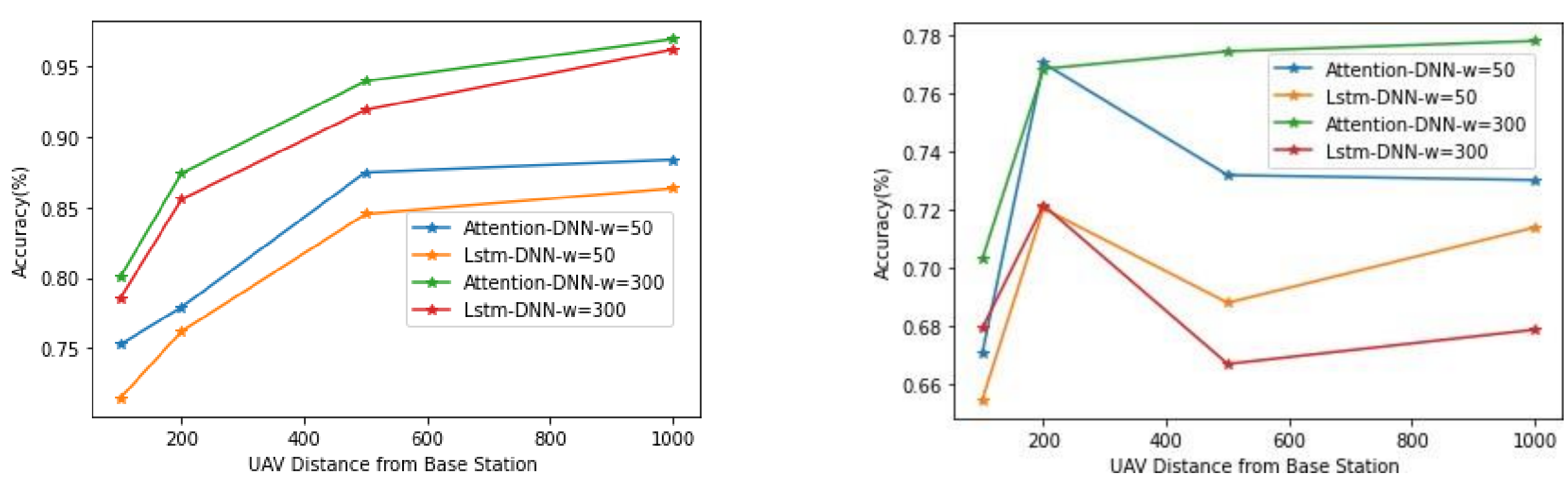

(a) LoS Condition (b) NLoS Condition

Fig. 5: Comparison between Attention and LSTM algorithms for w=50 and w=300, users =20, number of attackers = 2 attacker power =5 dBm (a) In LoS, (b) In NLoS

Table VII shows the differences in the overall accuracy between the Attention and LSTM layers for different window sizes (ranging from w=50 to w=300), different channel conditions (LoS, NLoS, and both) and the three proposed methods (DNN, DNN+Method1, DNN+Method2). Results are compared to the reference XGB algorithm, for different window sizes and channel conditions available in Table VIII. The XGB performs poorly when the hybrid dataset is applied to the algorithm in contrast to the results obtained with the DNN and DNN with Methods 1 and 2. Considering the same condition, better results are seen in the Attention layer except in DNN+Method 1 in NLos condition and w=100 where the difference is around 0.05 in favor of LSTM.

Moreover, we notice that an increase in the window size has a positive impact in the overall accuracy when using Attention layers, but for LSTM in NLoS conditions it has the opposite effect when w > 100. The pattern recognition in NLoS is hard to be extracted in general due to the low power received in the authenticated UAV, but for this particular case when w > 100, it decreases the overall accuracy.

With respect to the LoS, NLoS and Both conditions, LoS presents the best accuracy because there is no decrease in the received power due to obstacles and objects between the authenticated UAV and the small cell. Therefore, the deep network could learn the attacker pattern even in cases with channel variations and more users in the network. The combined condition presents the second-best results and as expected, NLoS presents the worst. Notice that by adding more nodes and layers, the deep network can learn this pattern, however there is a tradeoff in terms of memory and energy consumption, which is not within the scope of this work. The greatest impact of the MVA and TSA in the DNN is in NLoS conditions. Method 1 increases the overall accuracy by more than 10% when using LSTM and by approximately 10% with the Attention. Among the methods in the study, the MH-DNN + Method 2 performs better for LoS, whereas the MH-DNN + Method 1 performs better for NLoS conditions.

Fig. 5 depicts the accuracy against the distance between the authenticated UAV and the small cell in the network for two different window sizes using Attention and LSTM layers, for (a) LoS and (b) NLoS channel conditions. For each condition, we present the results for MH-DNN with no additional methods. Fig. 5(a) shows that, for LoS, both Attention and LSTM configurations with window size 300 (w=300) outperform the configurations with window size 50 (w = 50). In NLoS condition, see Fig. 5(b),the DNN embedded with the Attention layer has the best performance independently of the window size.

### C. *Comparison with other machine learning classifiers*

Fig. 6 compares the proposed DAtR (composed by MH-DNN, Method 1 and Method 2), with three other machine learning methods, namely RF, CAT, XGB, over the distances between the small cell and the authenticated UAV available in the dataset, in LoS and NLoS conditions, separately.

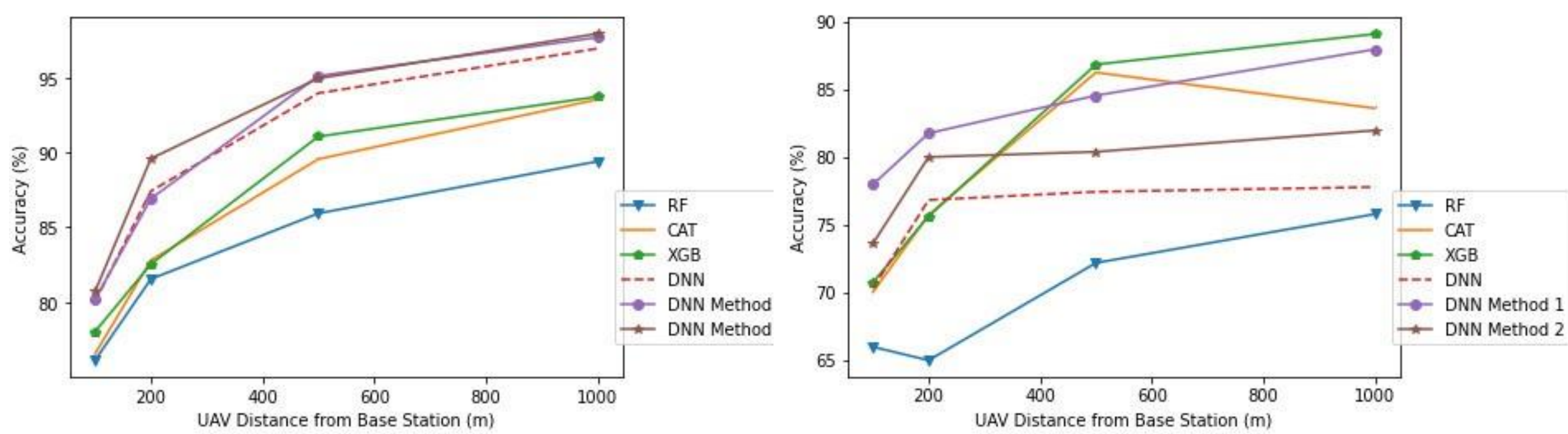

(a) LoS Condition (b) NLoS Condition

Fig. 6: Comparison between the proposed DNN with DNN+ Method 1, DNN + Method 1 and 2, RF, CAT, XGB. w=300, users =20, number of attackers = 2, attacker power =5 dBm. (a) In LoS, (b) In NLoS

We eliminate GNB and LR from the charts because they fail to achieve 70% accuracy across the range of distances and SVM because its performance is comparable to the other ML algorithms for shorter distances but dropped to 75% accuracy for those with d > 200 in LoS conditions. In Fig. 6 (a) we show that even our worst classifier, which is the MH-DNN embedded alone with the Attention layer, consistently outperforms well known classifiers such as RF, CAT, and XGB, while Method 1 and 2 present an additional improvement, especially for large distances. CAT and XGB perform similarly, while RF decreases its overall accuracy for large distances.

In general, compared to all the accuracies obtained from other algorithms, the proposed DAtR achieves an accuracy range from 80% up to 95% over all distance ranges. The mean accuracy the DatR achieves is 89.97%, while the RF, CAT and XGB achieved 83.24%, 85.60%, and 86.33% respectively.

Fig. 6 (b) presents the results for the NLoS channel condition. This figure shows that Method 1 in this case is more effective in short distances. However, note that the DAtR and Method 2 outperform the benchmark schemes for short distances, but they lose accuracy for higher distances. As such, Method 1 appears to achieve a good compromise between small and large distances.

Overall, comparing both charts, it is clear that DAtR can more easily identify attackers in LoS, but it can also be implemented in NLoS, or in mixed conditions depending on the link distance.

### D. Confusion matrices

Fig. 7 (a) and (b) illustrate the confusion matrices resulting from the proposed algorithms: MH-DNN, MH-DNN+Method 1 and ML alg, and MH-DNN+Method 2 and ML alg, for LoS and NLoS, respectively. We utilize the XGB as a ML algorithm for Method 1 and 2.

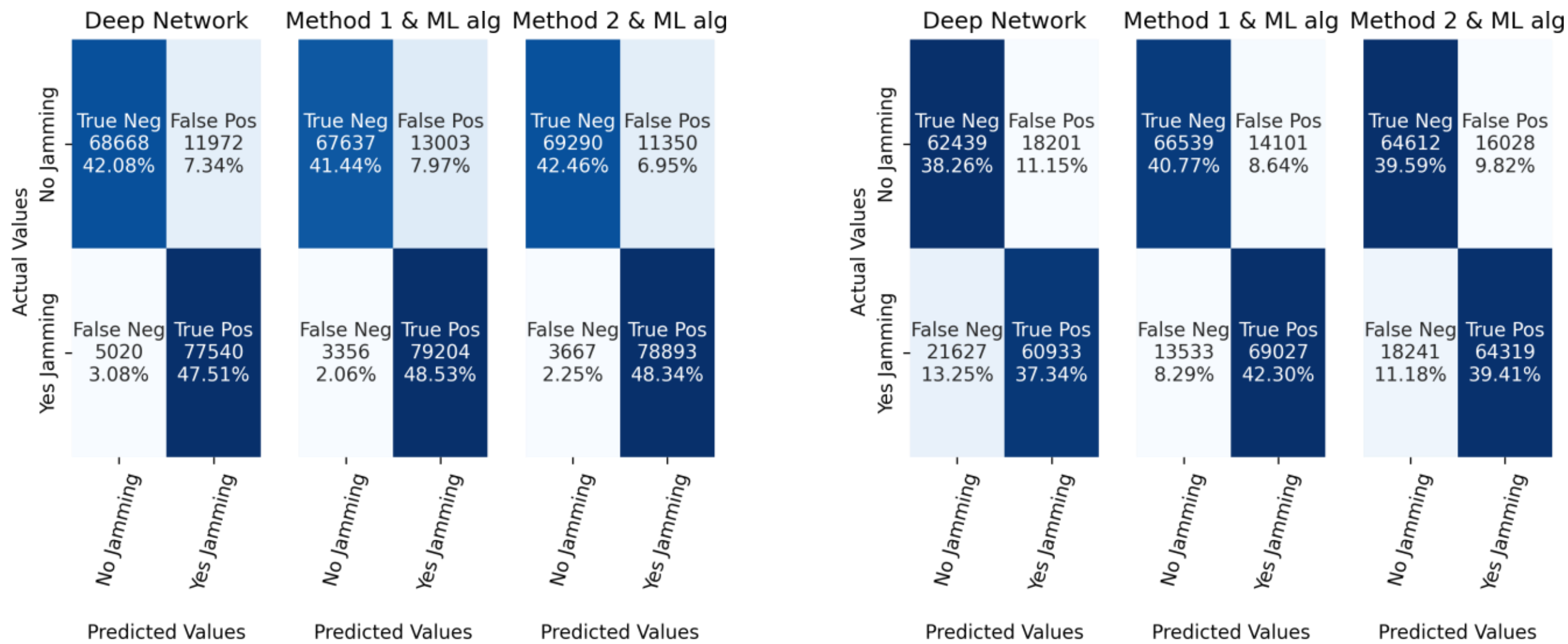

(a) LoS Condition

(b) NLoS Condition

Fig. 7: Overall Confusion Matrices of the proposed MH-DNN, MH-DNN+Method 1 and ML alg, and MH-DNN+Method 2 and ML alg, w=300, (a) In LoS, (b) In NLoS

We compare the results of Method 1 and Method 2 with the results of MH-DNN alone. We notice that MH-DNN+Method 2+XGB increases the accuracy in LoS scenarios. While MH-DNN+Method 1+XGB is more suitable for NLoS settings. For example, Fig. 7a highlights the difference between the two True Negative (`True Neg`) when we subtract Method 1 and Method 2 values from the Deep Network (MH-DNN).

Method 1+XGB results in -0.64% less accuracy, while with Method 2+XGB there is +0.38% better accuracy. Also, Method 1 increases the chances of False positive (`False Pos`) by +0.63 % while Method 2 decreases the likelihood of `False Pos` by -0.39 %.

In Fig. 7b, we see the opposite effect. Method 1+XGB has better values for `True Neg` and `False Pos` than Method 2 +XGB when comparing both to the Deep Network. When it comes to LoS, the MH-DNN+Method 2 performs better than the other approaches in the research, but the MH-DNN+Method 1 is the clear winner when it comes to NLoS.

Taking into account the best outcomes that we have so far, specifically, MH-DNN configured with Attention + Method 2 for LoS or + Method 1 for NLoS and XGB algorithm, except when explicitly mentioned, we use this configuration to show detailed performance evaluation considering all cases and parameters available in the dataset using DAtR. In the combined condition, we used MH-DNN configured with Attention + Method 1 for NLoS and XGB algorithm.

The accuracy presented in the confusion matrix, is the average accuracy from all the scenarios in the dataset, and it has a significant impact in the specific cases, as it will be presented in the next sections.

### E. Attacker number and power

Fig. 8 presents the accuracy over the number of attackers and their power, in (a) LoS, (b) Combined and (c) NLoS conditions. If we take close look at the individual charts, we see that the accuracy increases with more attackers and more power for LoS and Combined conditions. In the NLoS case, the low accuracy is centered in the scenario with 2 attackers when both of them are configured with power less than 5 dBm. It increases for both more and less attackers and as the attacker power rises.

In the LoS case, the scenario with 1 attacker is the hardest for the proposed algorithms to learn. In the Combined condition, 0 and 1 attacker scenarios are complicated for the algorithms to learn and for the NLoS condition, the most complicated scenario is with 2 attackers. In Los and Combined cases, the changes in the power presented improvements in the accuracy of around 5%. The low accuracy when there are less then 3 attackers in the scenario might be justified by the stochastic channel models available in 5G UAV cases where the channel adjustments experienced by the UAV can change approximately 30dB from one channel update to another. The number of users affects the total received power reducing the DAtR's overall accuracy. In the NLoS case, the fact that there are no straight rays feeding in the receiver impacts the overall power received and decreases the accuracy results. Comparing all the results, the NLoS simulation presents the lowest overall accuracy from all conditions, but the best accuracy that it can achieve is 93% with 4 attackers configured with 20 dBm power.

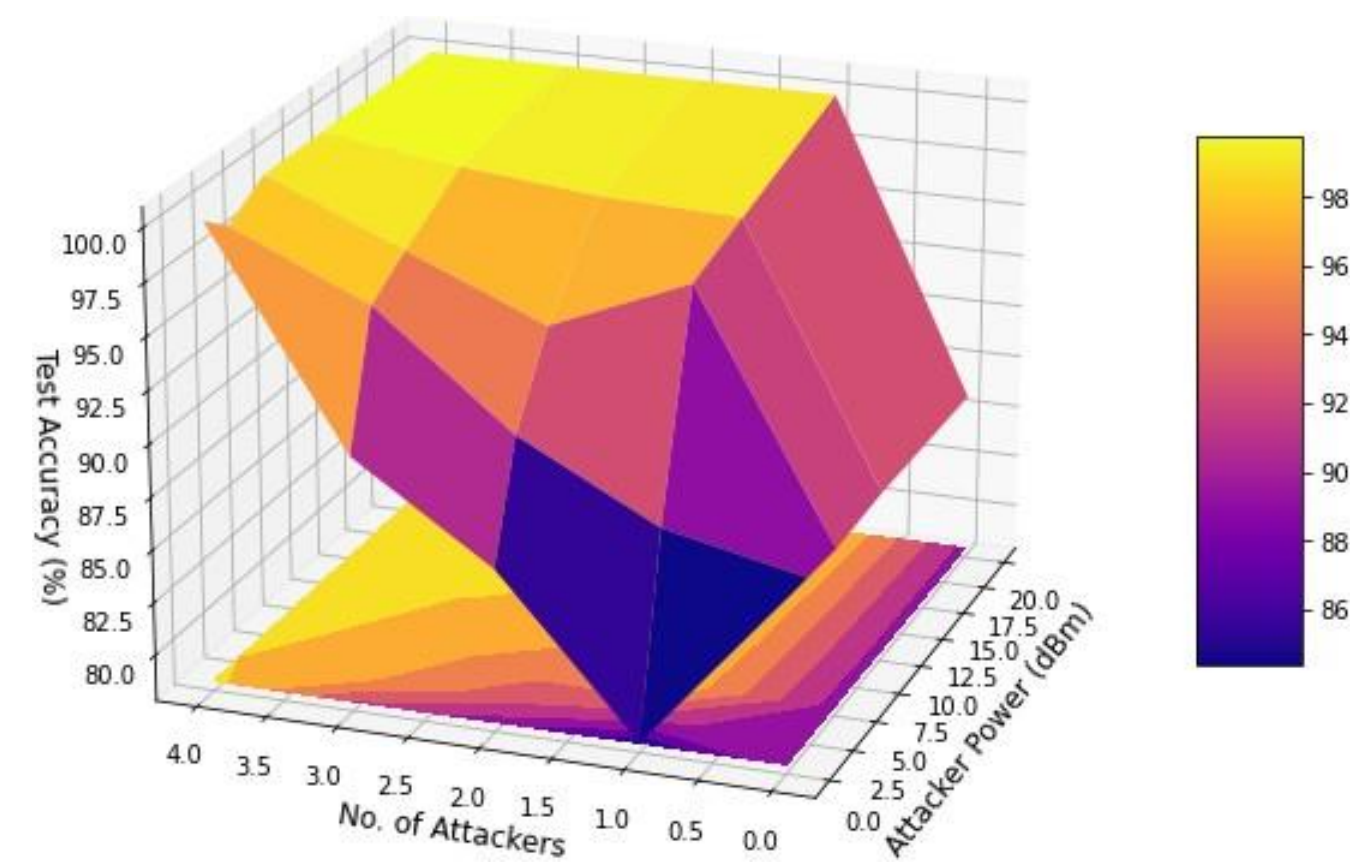

(a) LoS Condition

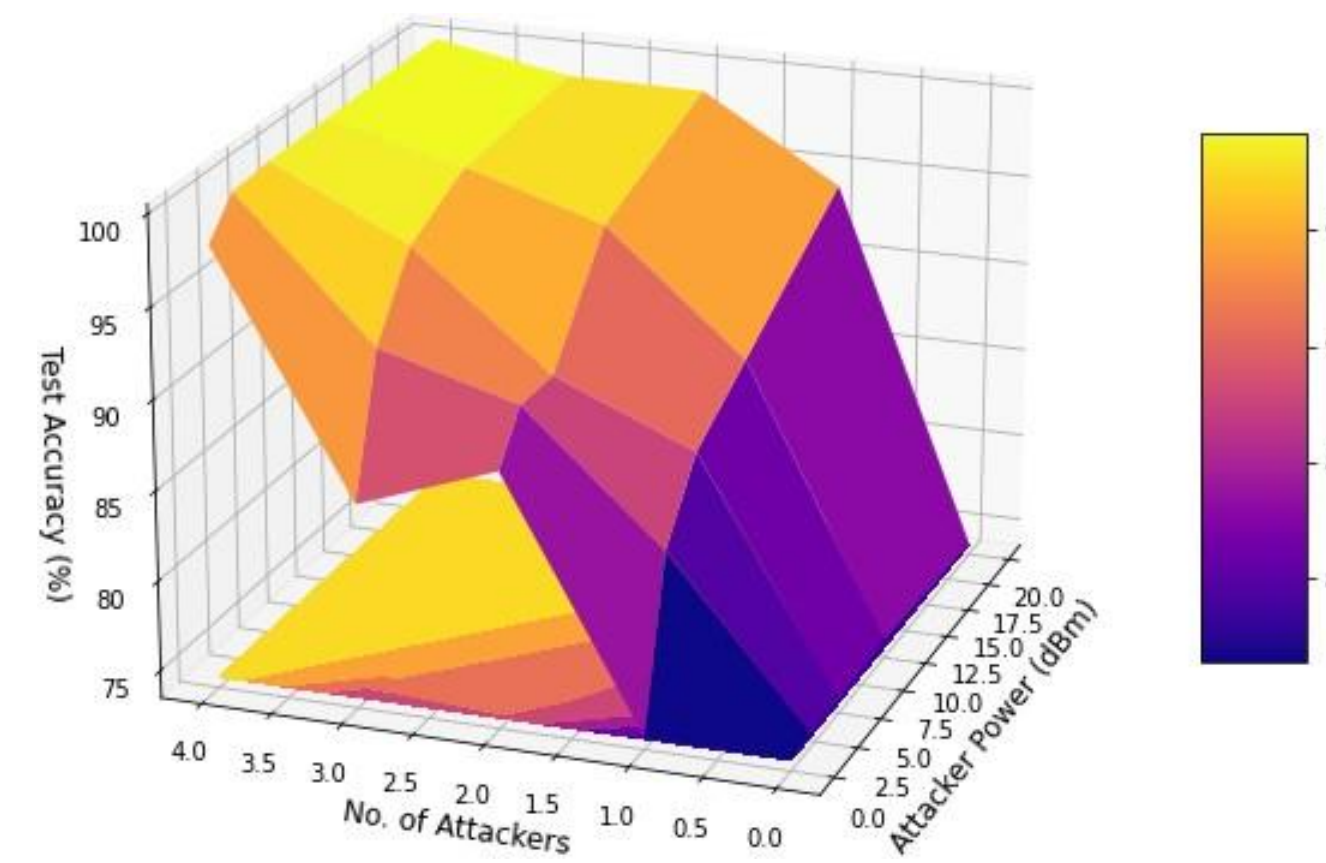

(b) Combined Condition

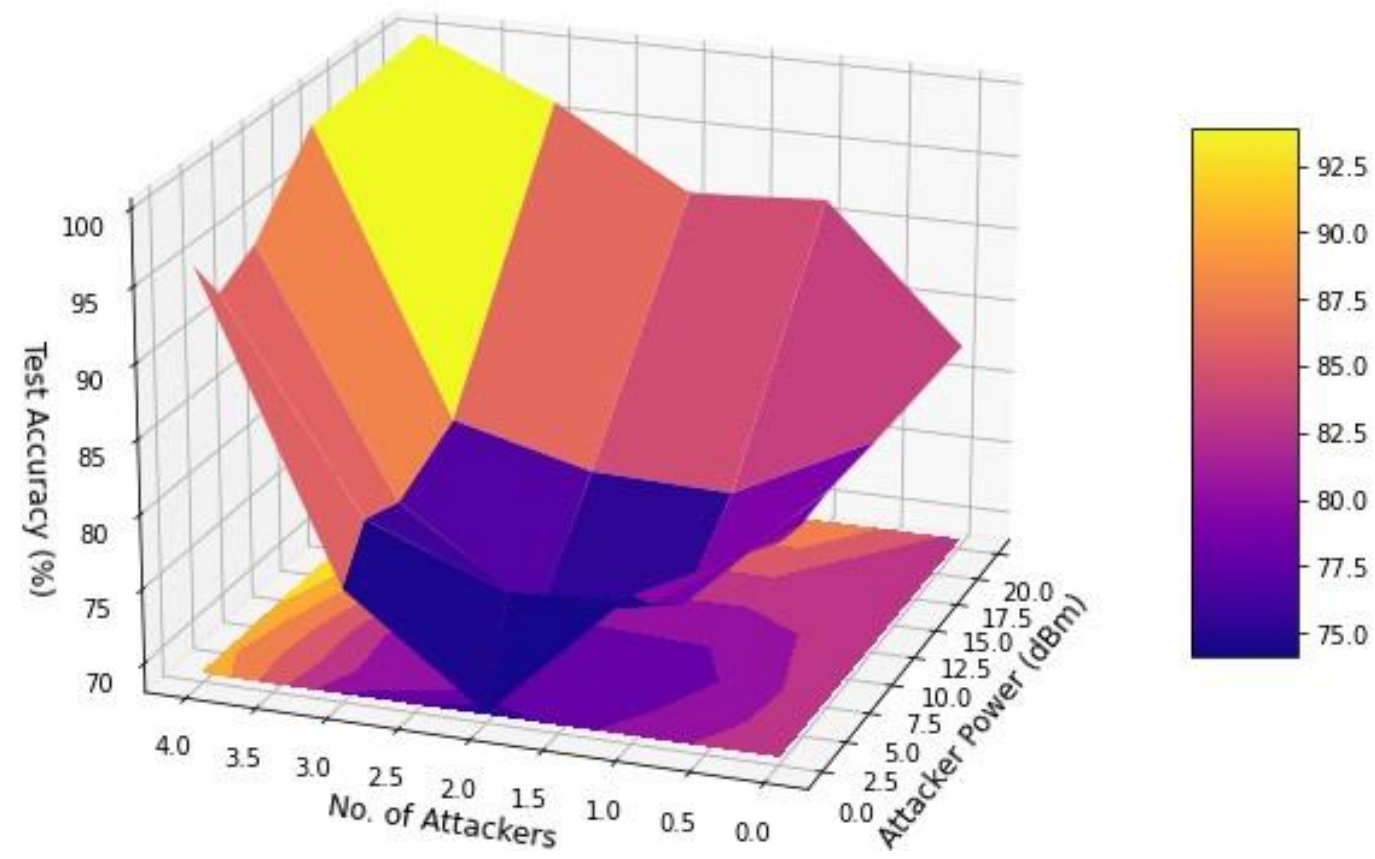

(c) NLoS Condition

Fig. 8: Accuracy vs Attackers Number and Attacker Power data during the test, users=20, distance=100, w=300, (a) LoS only, (b) LoS and NLoS, (c) NLoS only.

### *F. Comparison with data that is not in the training*

Fig. 9 (a) and (b) depict the accuracy results based on the attacker power when the number of users in the network is U=20, for distance = 500 and number of attackers = 2. We remove the data related to the attacker power = 2 and 10 (dBm) from the training. Therefore, the deep network sees both these pieces of data for the first-time during testing. We executed this simulation for LoS and NLoS conditions.

Fig. 9a demonstrates the outcomes for LoS. We notice a proportional decrease in all samples when we compare when the training is done with all and removed samples. This difference is around 1.5%. For the NLoS case, illustrated in fig 9b, there is a difference more significant than 0.5% only when the attacker was set up with 20 dBm power. There were no significant differences for the other cases, which shows the robustness of our proposed algorithm.

### *G. Attacker power and distance*

Fig. 10 shows the accuracy over distance and attackers power ratios during training for the three conditions: LoS, Combined, and NLos. In the three conditions, attackers with lower power are harder for the deep network to recognize.

In the LoS conditions, the deep network can identify attacks even though the base station is positioned 1000 m away from the authenticated UAV and the attacker power is lower than 5 dBm with 96% accuracy. There are improvements when the power increases, but we achieve better results when increasing distance. We believe that the interference from the users decrease at this position and that is why the deep network could achieve high accuracy. In the Combined condition, we see the impact of power on accuracy more clearly than in LoS. For example, when the attacker power is set to 15 dBm the accuracy is 85% when the distance between the authenticated UAV and the Base station is 100 m. We see a peak accuracy when the distance is 500m and the attacker power is 15dBm. While it is easier to identify attackers for the other conditions when the attacker power is higher than 5 dBm, in the NLoS condition, the attacker power needs to be adjusted to 15dBm so the deep network can have approximately 84% accuracy.

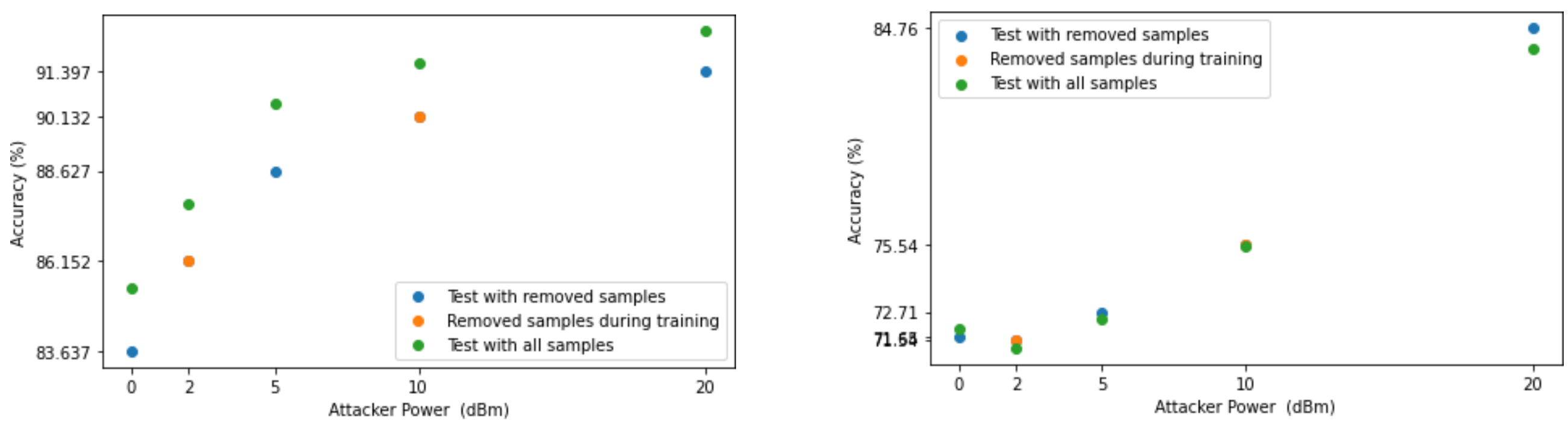

(a) LoS Condition

(b) NLoS Condition

Fig. 9: Comparison with data that is not in the training (a) In LoS, (b) In NLoS

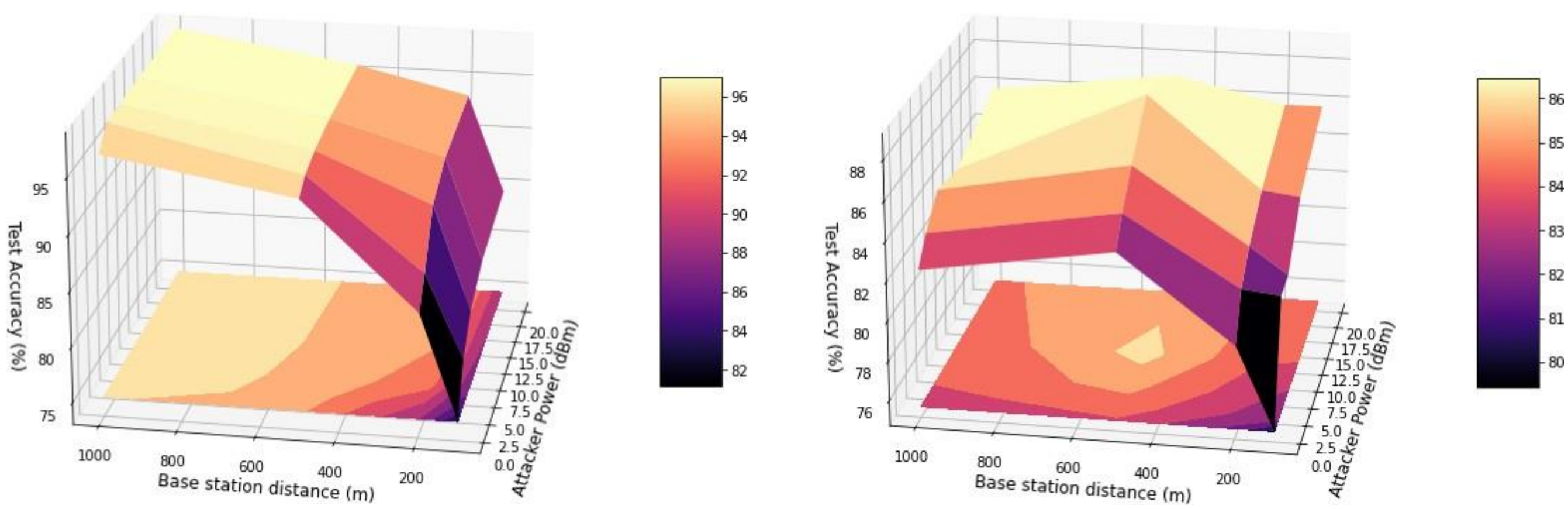

(a) LoS Condition

(b) Both Condition

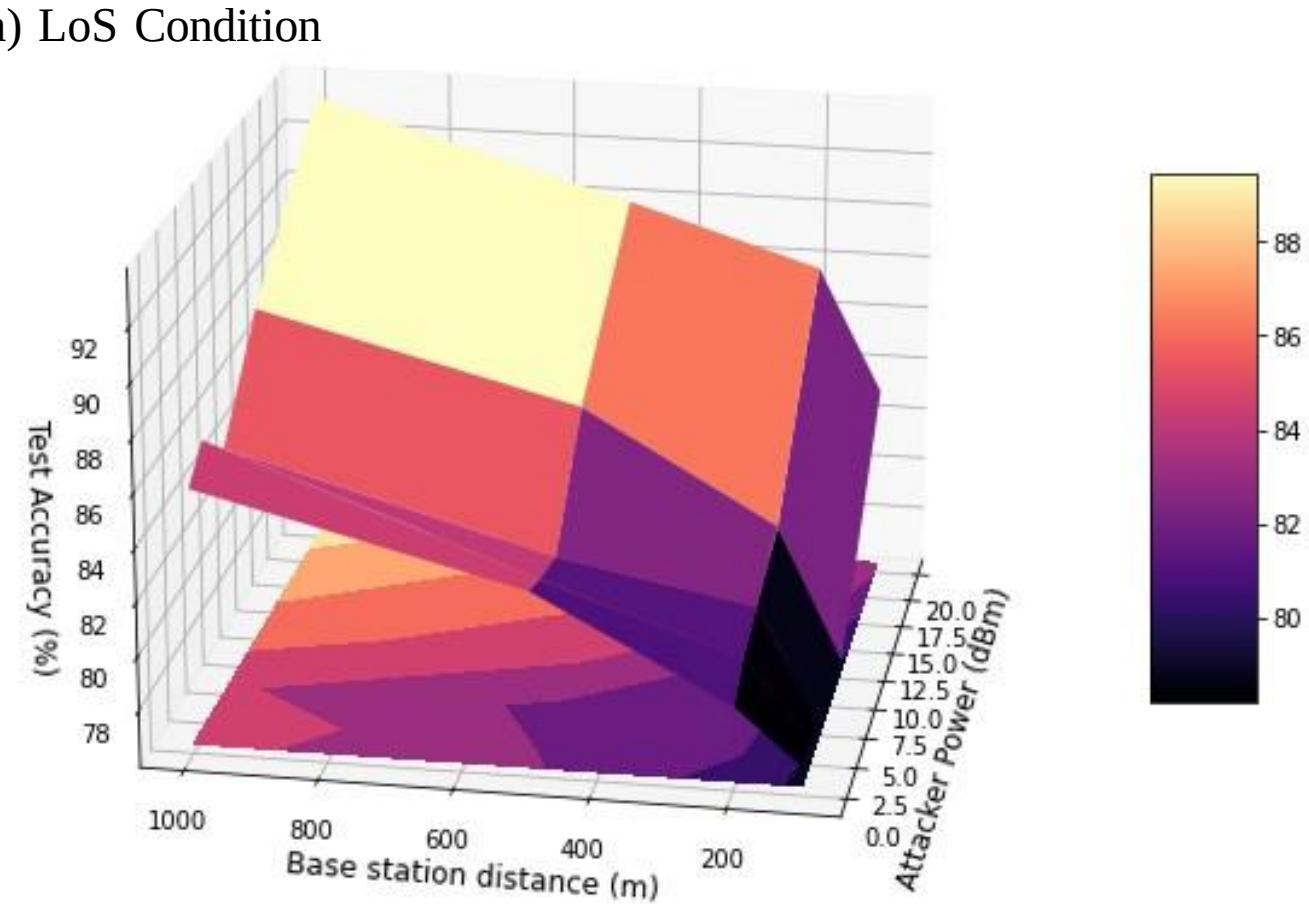

(c) NLoS Condition

Fig. 10: Accuracy vs Attackers Power and Attacker Distance test data, window_size=300, number of attackers=2, users =20 (a) LOS only , (b) LOS and NLoS , (c) NLoS only.

### *H. Average processing time*

Fig. 11 compares the average prediction time after training for each of the three baseline classifiers (RF, CAT and XGB) and the proposed MH-DNN configured with Attention or LSTM for different window sizes to classify each sample. Table IX shows the average values with their respective standard deviations. The prediction time is an important metric because it shows the latency in discovering attacks when using such algorithms in the UAVs. All timing tests were done using a system with a Nvidia RTX 3090 GPU.

TABLE IX: Prediction timing versus window size for the proposed deep network and three other ML classifiers

| | w=50 | w=100 | w=200 | w=300 |
|---|---|---|---|---|
| DNN-Attention | 30.9 ms±248$\mu s$ | 30.9 ms±335$\mu s$ | 31.9 ms±656$\mu s$ | 30.8 ms±391$\mu s$ |
| DNN-LSTM | 31.3 ms±1.03$ms$ | 31 ms±351$\mu s$ | 31.2 ms±311$\mu s$ | 30.5 $ms \pm 393 \mu s$ |
| CAT | 0.517 ms±561$ns$ | 0.824 ms±744$ns$ | 1.49 ms±939$ns$ | 2.19 ms±2.02$\mu s$ |
| RF | 71.6 ms±1.28$ms$ | 74.8 ms±1.63$ms$ | 76.6 ms±1.66$ms$ | 79.4 ms±1.76$ms$ |
| XGB | 0.657 ms±22.4$\mu s$ | 0.665 ms±22.5$\mu s$ | 0.683 ms±23.9$\mu s$ | 0.741 ms±21.6$\mu s$ |

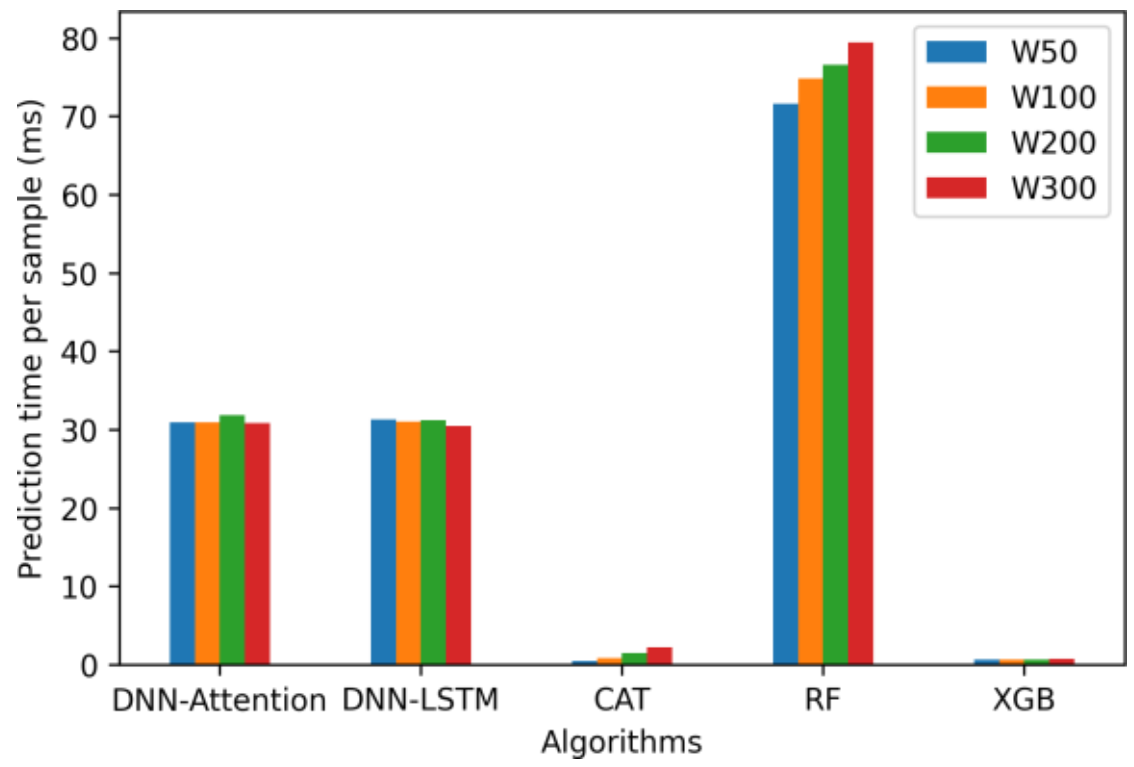

Fig. 11: Average processing time for each classifier

In Fig. 11, we can see that the window size has a small effect for the XGB and for the MH-DNN configured with Attention ot LSTM. However, it has a bigger impact on CAT and RF. The prediction time for CAT increases four times when the window size is 300. For RF, the impact of the window size is smaller than for CAT, but it still increases approximately 10% for the same window size (w=300). There is a minor difference between the LSTM and Attention prediction times. The RF algorithm displays the highest prediction time. Our proposed method has a good tradeoff between accuracy and prediction time.

## VI. Conclusion

This paper studied the attacks self-identifying problem in 5G UAV networks assuming scenarios with LoS, NLoS and a probabilistic combination of both conditions. Specifically, we proposed a small deep network system, named DAtR, that can cope with the attack self-identifying problem, and we verified its accuracy through extensive simulation campaigns. Our research examined five major implementation issues related to the deep network: how the key parameters, such as the window size, impact the deep network accuracy, the impact of different layers on the overall performance (i.e., Attention vs. LSTM), its performance compared to other machine learning alternatives for classification, the robustness of our deep network using data that is not available in training, and the prediction timing for the proposed DAtR. We showed that the proposed system, compared to six popular classifiers available in the literature, is a competitive option for the attack classification for all distance ranges in LoS conditions and for short range distances in NLoS conditions. The comparison between LSTM and Attention shows that increasing the window size in the LSTM setup reduced the performance, while with Attention it boosted performance. The use of Attention layers in DAtR outperformed the same system configured with LSTM. Finally, we present the performance graphs we created for each case study. Results have have demonstrated that our deep network reliably identifies attacks across all possible configurations. It was simpler to identify attacks in simulations with three or more attackers, fewer users, and a power of 10 dBm or higher. The identification accuracy was also affected by the three-dimensional distance between the small cell and the authenticated UAV. Here, the chances of identification improved with increasing distances since there was less interference to contend with.